\documentclass[aps, prd, twocolumn, lengthcheck, superscriptaddress, showpacs, nofootinbib]{revtex4-1}

\usepackage{times}
\usepackage{amsmath,amssymb,amsfonts}
\usepackage{bm}
\usepackage{graphicx}
\DeclareGraphicsExtensions{.pdf,.eps,.png,.jpg}
\usepackage{acronym}
\usepackage{color}
\usepackage{xcolor}
\usepackage[caption=false]{subfig}
\usepackage{tabularx}
\usepackage{longtable}
\usepackage{multirow}
\usepackage{dcolumn}
\usepackage{soul}
\usepackage{float}
\usepackage{hyperref}
\usepackage{cleveref}

\allowdisplaybreaks

\def\roughly#1{\mathrel{\raise.3ex\hbox{$#1$\kern-.75em%
\lower1ex\hbox{$\sim$}}}}
\def\lsim{\roughly<}

\begin{document}

\title{Ranking Candidate Signals with Machine Learning in Low-Latency Search for Gravitational-Waves from Compact Binary Mergers}
\author{Kyungmin~\surname{Kim}}
\email{kkim@kasi.re.kr}
\affiliation{Korea Astronomy and Space Science Institute, 776 Daedeokdae-ro, Yuseong-gu, Daejeon 34055, Republic of Korea}
\affiliation{Department of Physics, The Chinese University of Hong Kong, Shatin, New Territories, Hong Kong}

\author{Tjonnie~G.~F.~\surname{Li}}
\affiliation{Department of Physics, The Chinese University of Hong Kong, Shatin, New Territories, Hong Kong}

\author{Rico~K.~L.~\surname{Lo}}
\affiliation{Department of Physics, The Chinese University of Hong Kong, Shatin, New Territories, Hong Kong}
\affiliation{LIGO Laboratory, California Institute of Technology, MS 100-36, Pasadena, California 91125, USA}

\author{Surabhi~\surname{Sachdev}}
\affiliation{LIGO Laboratory, California Institute of Technology, MS 100-36, Pasadena, California 91125, USA}
\affiliation{Department of Physics, Pennsylvania State University, University Park, PA 16802, USA}

\author{Robin~S.~H.~\surname{Yuen}}
\affiliation{Department of Physics, The Chinese University of Hong Kong, Shatin, New Territories, Hong Kong}

\date[]{Last modified: \today}

\begin{abstract}
In the multi-messenger astronomy era, accurate sky localization and low latency time of gravitational-wave (GW) searches are keys in triggering successful follow-up observations on the electromagnetic counterpart of GW signals. We, in this work, focus on the latency time and study the feasibility of adopting supervised machine learning (ML) method for ranking candidate GW events. We consider two popular ML methods, random forest and neural networks. We observe that the evaluation time of both methods takes tens of milliseconds for $\sim$ 45,000 evaluation samples. We compare the classification efficiency between the two ML methods and a conventional low-latency search method with respect to the true positive rate at given false positive rate. The comparison shows that about 10\% improved efficiency can be achieved at lower false positive rate $\sim 2 \times 10^{-5}$ with both ML methods. We also present that the search sensitivity can be enhanced by about 18\% at $\sim 10^{-11}$Hz false alarm rate. We conclude that adopting ML methods for ranking candidate GW events is a prospective approach to yield low latency and high efficiency in searches for GW signals from compact binary mergers.
\end{abstract}

\pacs{95.85.Sz, 98.70.Rz, 07.05.Mh}

\keywords{}

\maketitle


\section{Introduction}

Recently, ground-based gravitational-wave (GW) observatories, LIGO~\cite{aLIGO} and Virgo~\cite{aVirgo} detected GW170817 \cite{GW170817:prl} in about 1.7 seconds advance the observation of a short GRB, GRB170817A~\cite{GRB170817A:gcn} which was identified by the Fermi Gamma-ray Burst Monitor (GBM)~\cite{fermi_gbm}. These coincident observations of both GW and short GRB became a monumental event for opening the era of multi-messenger astronomy~\cite{mma:2017apjl}. From the joint observation, one of the most plausible scenarios for the central engine which powers a short GRB is confirmed too.

With the opening of multi-messenger astronomy era, it is natural to believe that we will observe other kinds of joint GW-electromagnetic (EM) events too as summarized in Ref.~\cite{Margutti:2018dpr} with future GW detectors and optical telescopes such as Large Synoptic Survey Telescope~\cite{Ivezic:2008fe}. For a joint GW-EM observation, we may use a GW event as a precursor for triggering follow-up observations on its EM counterpart. The success of this kind of joint observation will strongly depend not only on reducing the error of sky localization in GW detection but also on curtailing the latency of GW search; precise sky localization is related to how many GW detectors in various geographical locations are online simultaneously while the latency of search is associated with the quality of GW data and analysis efficiency. Here, the efficiency implies the accuracy of analysis. However, increasing the number of GW detectors is not trivial despite KAGRA~\cite{Somiya:2011np} and LIGO-India~\cite{LIGO_India} will come online in the near future in addition to currently operating LIGO and Virgo detectors. Improving the quality of GW data faces another difficulty because the current instrumental specifications are adopting state-of-the-art technology already. On the other hand, enhancing the efficiency of data analysis is relatively capable since studying the capability of a new method is much easier than others. Therefore, we focus on the analysis efficiency in this work.

Up to date, several pipelines~\cite{gstlal_inspiral:2017prd,spiir:2012prd,mbta:2012aa,pycbc:2016cqg,MBTA:2016cqg,cWB:cqg2008} for the low-latency GW search have been developed and conducted to search GW signals by analyzing the time series GW data in real-time. The common goal of these pipelines is identifying a candidate GW event as soon as possible. Currently, the latency between the actual event time and the identification of a candidate event with those search pipelines takes about a few minutes as reported in Ref.~\cite{mma:2017apjl} for the detection of GW170817. When a low-latency search pipeline successes in the identification of a candidate event based on the significance of a ranking method of each pipeline and obtains the information, e.g., event time, sky location (right ascension and declination), and signal-to-noise ratio of the candidate event, it forwards those information to a database system, GraceDB~\cite{gracedb}. Then GraceDB delivers those information to EM partner observatories/telescopes through an alert system such as Gamma-ray Coordinates Network~\cite{GCN} alert to trigger follow-up observations for seeking correlated EM events.

Meanwhile, a candidate GW event is a survived one from multiple stages of sanity tests of a search pipeline and the event contains the result of each sanity test too in addition to the observational information forwarded to GraceDB. Thus, we can regard identifying the origin of a candidate event as a \emph{multivariate classification problem} and the information describing candidate GW events seamlessly leads the consideration of machine learning (ML). Indeed ML has been gradually implemented and accepted in various GW data analyses~\cite{Cannon:2008cqg,Biswas:2013prd,Adams:2013prd,Rampone:2013,kim:2015cqg,Zevin:2017cqg,rfgw:2017prd,Mukund:2016thr} to achieve efficient, that is, accurate analysis not only for the identification of GW signals but also for the characterization of non-Gaussian transient noises. From these studies it has been shown that we can consider ML as an alternative and complementary method to the conventional ranking method of each analysis based on their classification performances. Hence, in this work, we study the feasibility of curtailing the latency of the low-latency search by adopting ML for ranking candidate events with maintaining high-efficiency.

This paper is organized as follows: we present brief descriptions on used tools, data preparation, and procedure of applying MLs in Sec.~\ref{sec:methods}. The result of classification performance of ML for the given data is summarized in Sec.~\ref{sec:performance}. In Sec.~\ref{sec:sensitivity}, we present the detection sensitivity obtained with the application of ML and compare it to the conventional ranking method, log-likelihood ratio, of the GstLAL inspiral search pipeline~\cite{gstlal_inspiral:2017prd}. Finally, in Sec.~\ref{sec:discussion}, we discuss the results of this work.


\section{Method}
\label{sec:methods}

We start with briefly introducing the tools used in this work.\footnote{Since describing details of used tools are out of the scope of this work, we recommend reader to refer references.} Then we describe the procedures from preparing data to obtaining the output of machine learning (ML).

\subsection{Tools}
\label{sec:tools}

\subsubsection{GstLAL Inspiral Search Pipeline}
GstLAL inspiral search pipeline~\cite{gstlal_inspiral:2017prd} (hereafter GstLAL pipeline) is designed for the low-latency search for gravitational-waves (GWs) radiated from compact binary mergers. It is built based on the GstLAL library \cite{gstlal} which was derived from the GStreamer \cite{gstreamer} and the LIGO Algorithm Library \cite{lal}. The pipeline produces candidate events from data of each GW detector by performing \emph{matched filtering}~\cite{Owen:1998dk} with template waveforms. In turn, if two or more detectors are online, the pipeline searches coincident events from detectors in network; given an event in one detector, the pipeline checks for corresponding events in the other detector within an relevant time window, which takes into account the maximum GW travel time between detectors and statistical uncertainty in the measured event time due to detector noise at the moment~\cite{gstlal_inspiral:2017prd}.

We can use the pipeline in two different modes, \emph{online} which makes low-latency identification of a candidate event and \emph{offline} which archives GW data with other information such as background statistics and data quality for further investigation on the candidate event identified from the \emph{online} mode. With the offline mode, in specific, it is possible to perform the \emph{software injection} -- injecting a bunch of simulated GW signals for compact binary systems into the calibrated GW data in order to test the search performance of the pipeline by comparing the physical parameters for the simulated signals and the recovered parameters by the pipeline.

Both modes of the GstLAL pipeline use \emph{log-likelihood ratio} \cite{far:2013prd} defined as
\begin{equation}
\ln{L}(\bm{\lambda}) = \ln \frac{P(\bm{\lambda} | s)}{P(\bm{\lambda} | n)} \label{eq:lnL_gstlal}
\end{equation}
as a ranking method to judge the significance of candidate events. In Eq.~(\ref{eq:lnL_gstlal}), $P(\bm{\lambda} | s)$ and $P(\bm{\lambda} | n)$ are the probability of observing parameters of $\bm{\lambda}$ of candidate events of all detectors given a GW signal, $s$, and background noise, $n$, respectively. The parameter vector $\bm{\lambda}$ consists of characteristic parameters of the candidate event such as signal-to-ratio, $\chi^2$, physical parameters of template waveforms used in identifying candidate events, detector sensitivities at the time of the event, mean trigger rates at the time of the event, the trigger phases, and inter-detector time differences (for details, see Refs.~\cite{far:2013prd, Sachdev:2019vvd}).

\subsubsection{Machine Learning}
We consider \emph{supervised} machine learning algorithms in this work since we will use \emph{prelabeled}\footnote{If one has \emph{unlabeled} data and wants to train a machine learning algorithm for either classification or regression, this kind of training is called \emph{unsupervised} learning.} data for training as described in the following subsection. Amongst many supervised learning algorithms, we adopt \emph{random forest} (RF)~\cite{Breiman:2001} and \emph{neural network} (NN)~\cite{McCulloch:1943}.

RF was suggested to remedy the biased overestimation problem of the classical \emph{decision tree} algorithm. RF is basically a collection of decision trees. However, it can reduce any biased effect in data classification by imposing (i) random selection in configuring input data for each tree and (ii) random choice on criteria at each binary split. Also, as an additional method for reducing biased overestimation problem, RF scores ranks on samples in the input data by averaging those ranks obtained from all decision trees in the forest.

NN operates as an artificial intelligence network similar with the biological neuron system: activation of a node is determined by an activation function which judges whether the strength -- sum of the value of each node times the value of connection from one node to another node in the next adjacent layer -- exceeds a certain criteria or not. 
Nowadays, NN can be divided into two categories, \emph{shallow} NN (SNN) and \emph{deep} NN (DNN), by the complexity of the structure of a network, more precisely, by the number of hidden layers: if there is one hidden layer, it is called as SNN and if there are two or more hidden layers, it is called as DNN. But it is known that if one can solve the issue on the computing time due to the complex structure of DNN, the performance of DNN is better than that of SNN in general.

In the implementation of these two MLs, we use two different open source packages: for RF, we use Scikit-Learn~\cite{scikit_learn} which supports various purposes of implementing machine learning algorithms such as supervised/unsupervised learnings or classification/regression problems. On the other hand, for NN, we use TensorFlow~\cite{tensorflow} because it allows constructing DNN efficiently by reducing computing time with sophisticated computational algorithms and/or multiple computing processors.

\subsection{Data Preparation}
\label{sec:data_preparation}

We use data obtained from LIGO Hanford, WA, U.S.A. (H1) and LIGO Livingston, LA, U.S.A. (L1). For the purpose of conducting classification with supervised MLs, we consider two classes of data, simulated signal data and background noise data. Hereafter we call the simulated signal data and background noise data as simply signal data and noise data for convenience. In particular, for the signal data, we consider binaries of black hole-black hole (BBH) because GWs from BBHs are the most common type of signal detectable by ground-based detectors.\footnote{In general, expecting electromagnetic (EM) counterparts for BBH events is mainly discussed in theoretical studies~\cite{Perna:2016jqh, Loeb:2016fzn, Stone:2016wzz, deMink:2017msu, McKernan:2017umu}. However a possible association of a gamma-ray burst to the first GW detection, GW150914 was discussed in the literature~\cite{Connaughton:2016umz}. Thus considering BBH signal in this work is viable about discussing latency with keeping in mind GW-EM joint observations.} Hence, we use mock data\footnote{To avoid overfitting that could be occurred in the training of MLs, we need sufficient number of signal samples, at least $> \mathcal{O}(10^3)$. For this reason, simulated mock data is favorable than a single real signal of GW150914.} of GW150914 generated by the software injection with the offline mode of GstLAL pipeline. Meanwhile, for the noise data, we use the data obtained by running \emph{time-slide} with GstLAL pipeline around the time of software injection. The software injection and time-slide were conducted with a chosen data segment taken between October 21, 2015 UTC and December 3, 2015 UTC, where no GW signal was found during the O1 operation of LIGO and Virgo.

From the coincident events of H1 and L1 data, we extract six feature parameters; signal-to-noise ratio (SNR) and chi-square statistic of each trigger as statistical feature parameters and, as physical feature parameters, masses and spin magnitudes of two component compact objects. We use the same feature parameters for the configuration of both signal and background data consistently.

We first shuffle the samples of the input data to reduce any biased effect in the composition of samples. Then, we divide the shuffled input data into two categories, train and test data, such as 75\% of the whole data for the train data and the rest 25\% for the test data. We use the train data and test data to train ML and to evaluate the performance of trained ML, respectively. The number of signal and noise samples for train and test data is tabulated in Table \ref{number_of_samples}.
\begin{table}[h!]
\caption{Number of signal and noise samples for training and test data.}
\begin{center}
\begin{tabularx}{\columnwidth}{X  c  X  r  X  r}
\hline
\hline
\multicolumn{2}{c}{} & & {Signal} & & {Noise} \\
\hline
\multirow{2}{*}{H1} & Train & & 3,641 & & 129,405 \\
\cline{2-6}
 & Test & & 1,220 & & 43,129 \\
\cline{1-6}
\multirow{2}{*}{L1} & Train & & 3,623 & & 129,423 \\
\cline{2-6}
 & Test & & 1,238 & & 43,111 \\
\hline
\hline
\end{tabularx}
\end{center}
\label{number_of_samples}
\end{table}
One can recognize that the number of signal and noise samples are imbalanced which may lead biased training. However, the imbalance may mimic the real situation since, in real detections, identifying a GW signal from noise dominant GW data is common. Thus we admit the imbalance and aim that successfully classifying desired signal samples from much larger number of noise samples as a challenge of this work.

\subsection{Training and Evaluation}
\label{sec:training}

We train each ML in different manner not only because of the different characteristics of tested MLs, RF and NN, but because of the different properties and usages of implemented packages, Scikit-Learn and TensorFlow. For given data, optimal choice on the hyperparameters of MLs in the training procedure is critically related to the performance of each ML. We determine the hyperparameters of RF and NN with the strategies described in Appendix~\ref{sec:adx_rf_hyp} and \ref{sec:adx_nn_hyp}, respectively, and use them to train each ML. Once the training is done, the trained ML is recalled for the evaluation of test data.

We evaluate the test data by using the trained MLs. At this stage, each ML scores a rank, $r$ on each sample of the test data based on the probabilistic prediction. Thus, the value of $r$ is given within a range of $0 \leq r \leq 1$.  We observe that the evaluation time for scoring ranks on about 45,000 samples in the test data takes about tens of milliseconds.

\begin{figure*}[ht!]
\subfloat[H1 data]{\label{fig_4-1} \includegraphics[width=0.8\columnwidth]{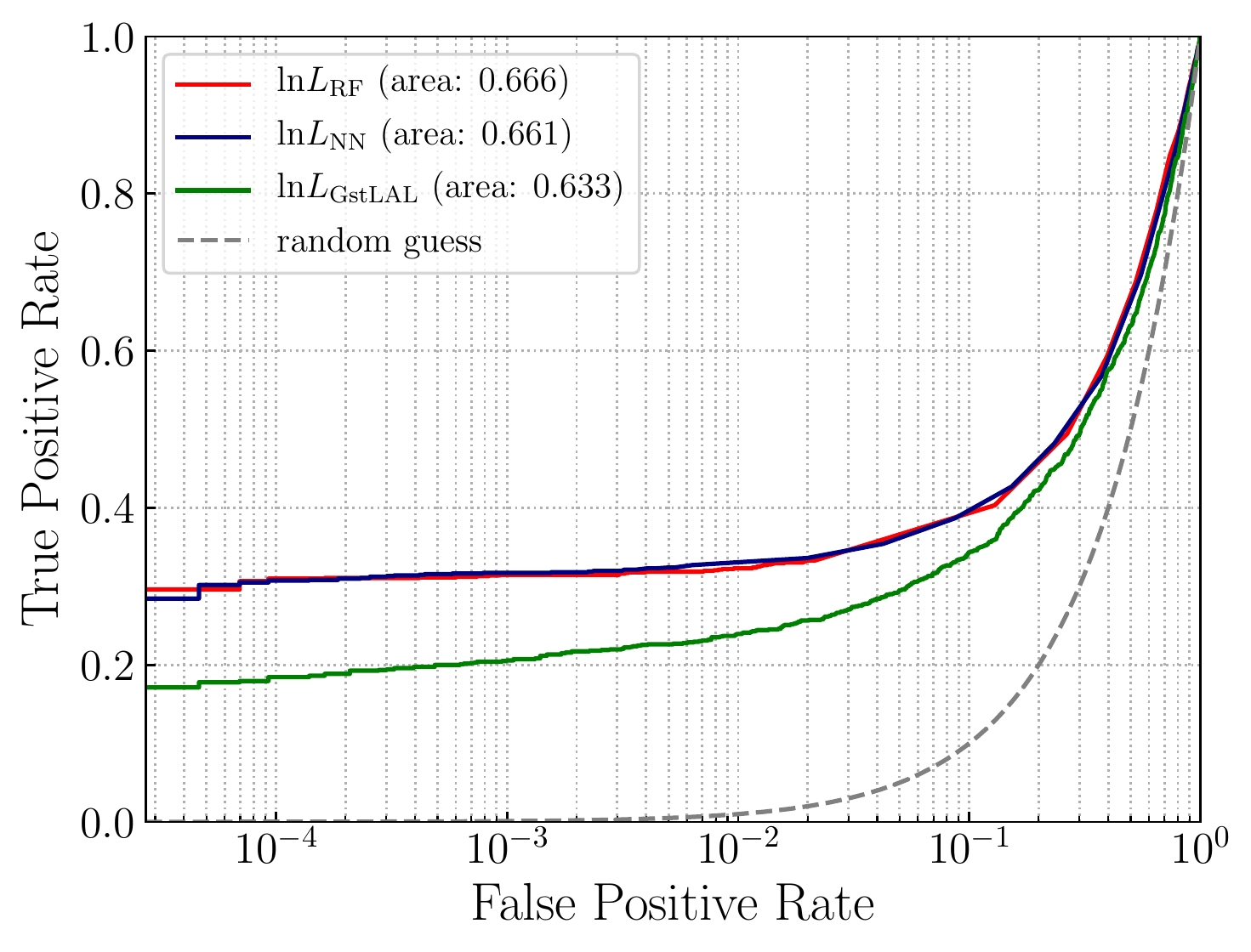}}
\subfloat[L1 data]{\label{fig_4-2} \includegraphics[width=0.8\columnwidth]{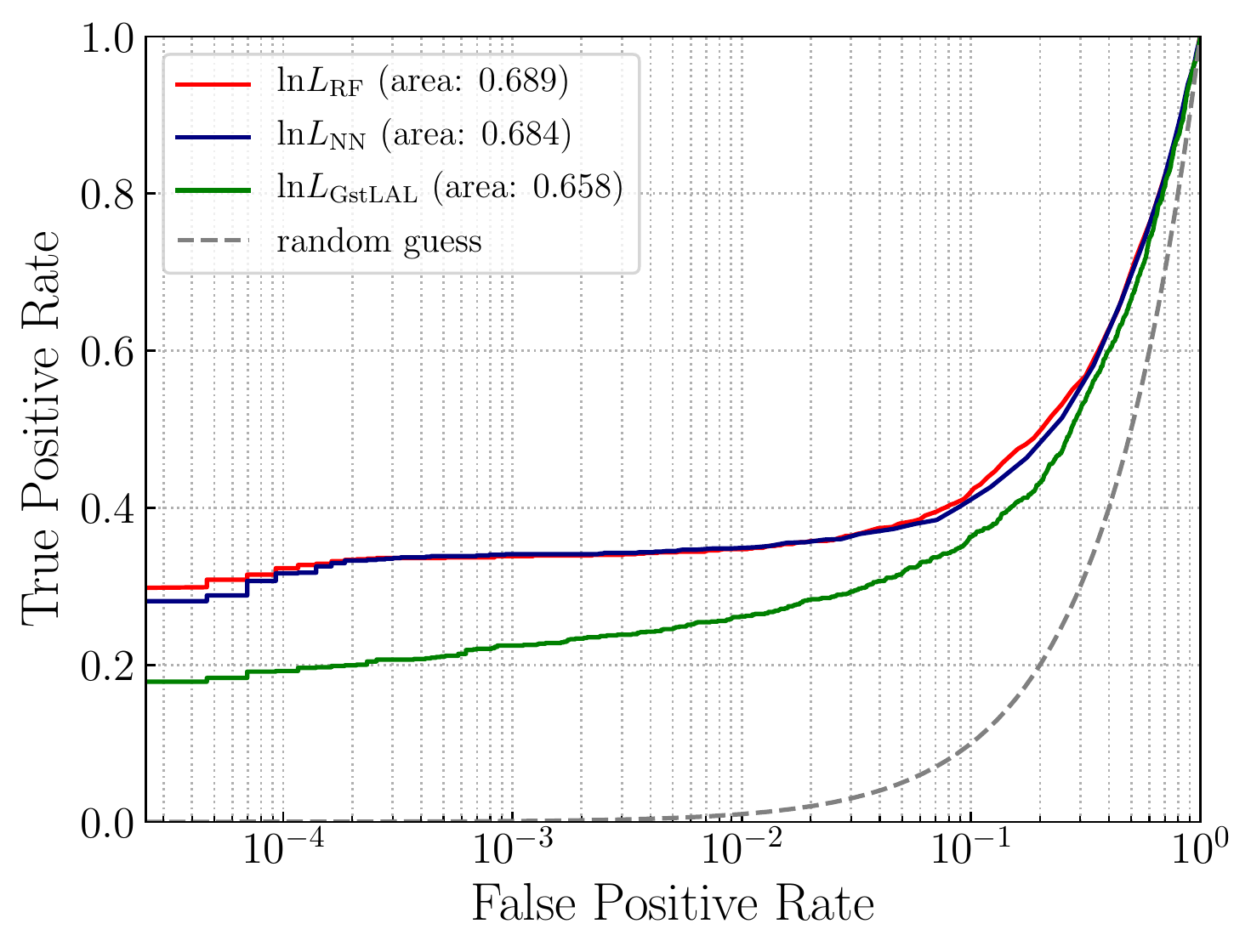}}
\caption{(Color online) Combined ROC curves and values of area under curve. The left and right panels show the ROC curves of the result of H1 or L1 data, respectively. The red and navy solid lines indicate the results of $\ln{L}$s of MLs and the green solid line indicates the result of $\ln L$ of GstLAL pipeline. One can see that both MLs show higher TPR than GstLAL pipeline over given FPR ranges and it results in that the area under curves of MLs are larger than the AUCs of $\ln L$. The performance of RF and NN are more or less similar to each other.}
\label{fig_roc}
\end{figure*}


\section{Classification Performance}
\label{sec:performance}

We discuss the classification performance of trained MLs in this section by comparing it to the performance of the ranking method of GstLAL pipeline. 

As described in the previous section, MLs return only probabilistic values between 0 and 1 while GstLAL returns unnormalized values of the log-likelihood ratio with Eq.~(\ref{eq:lnL_gstlal}). To address this issue, we compute log-likelihood ratio with the resulted ranks obtained from the evaluation such as
\begin{equation}
\ln{L}(r) = \ln \frac{P(r | s)}{P(r | n)} \label{eq:lnL_ML}
\end{equation}
by following the same analogy of Eq.~(\ref{eq:lnL_gstlal}) because (i) we can separate samples of the test data into either $s$ or $n$ based on the prelabeled class and (ii) we can estimate the probability density function of the given ranks of signal and noise samples too. Thus, by these two reasons, this approach is applicable to the evaluated result too and this prescription makes the comparison to be fair. 

The estimation of probability densities of the numerator and denominator of Eq.~(\ref{eq:lnL_ML}) is done by using the \emph{Kernel Density Estimation} method of Scikit-Learn with Gaussian kernel and an empirically determined optimal bandwidth of $0.03$. One can find the result of probability density estimation from Appendix~\ref{sec:adx_pdf}.

\subsection{ROC Curve}
\label{sec:roc}

In order to discuss the performance, we draw the receiver operating characteristic (ROC) curve as a figure-of-merit. To draw the ROC curve, we define \emph{true positive rate} (TPR) and \emph{false positive rate} (FPR) as follows
\begin{eqnarray}
\textrm{TPR} &\equiv& \frac{N^{(s)}(\ln{L}^{(s)}_{}( \ln{L} \geq \ln{L}_\textrm{th} ))}{N^{(s)}_\mathrm{T}} \nonumber \\
&\equiv& P(\ln{L}^{(s)} ( \ln{L} \geq \ln{L}_{\textrm{th}})), \label{eq_tpr} \\
\textrm{FPR} &\equiv& \frac{N^{(n)}(\ln{L}^{(n)}_{}( \ln{L} \geq \ln{L}_\textrm{th} ))}{N^{(n)}_\mathrm{T}} \nonumber \\
&\equiv& P(\ln{L}^{(n)} ( \ln{L} \geq \ln{L}_{\textrm{th}})), \label{eq_fpr}
\end{eqnarray}
where $N^{(s)}$ and $N^{(n)}$ respectively denote the number of signal and noise samples satisfying their values of $\ln L$ are larger than or equal to a given threshold value, $\ln L_\mathrm{th}$ within the group of signal samples, $\ln{L}^{(s)} = \{ \ln{L}_i; i=1,2,...,N^{(s)}_\mathrm{T} \}$ and the group of noise samples, $\ln{L}^{(n)} = \{ \ln{L}_j; j=1,2,...,N^{(n)}_\mathrm{T} \}$. Therefore, $\ln{L}^{(s)}(\ln{L} \geq \ln{L}_\textrm{th})$ or $\ln{L}^{(n)}(\ln{L} \geq \ln{L}_\textrm{th})$ represent subgroups of $\ln{L}^{(s)}$ or $\ln{L}^{(n)}$, respectively, satisfying $\ln L \geq \ln L_\mathrm{th}$. $N^{(s)}_\mathrm{T}$ and $N^{(n)}_\mathrm{T}$ are respectively the total number of signal and noise samples of test data presented in Table~\ref{number_of_samples}. Note that TPR and FPR represents how likely identifying signal sample as signal correctly and noise sample as signal incorrectly respectively. Hence, we desire to obtain higher value of TPR than FPR as $\ln L_\mathrm{th}$ increases. Subsequently, we can interpret a ranking method resulting higher TPR at lower FPR as better discriminator in distinguishing signal from noise adequately.

We present ROC curves in Fig.~\ref{fig_roc} by computing TPRs and FPRs with Eqs.~(\ref{eq_tpr})~and~(\ref{eq_fpr}) respectively. To depict the tendency of TPR with respect to FPR, we limit the the range of $\ln L_\mathrm{th}$ as $\ln{L}^{(n)}_\textrm{min} \leq \ln{L}_\textrm{th} \leq \ln{L}^{(n)}_\textrm{max}$, i.e., to make the minimum FPR to be $1/N^{(n)}_\mathrm{T}$. 
In the legend box of Fig.~\ref{fig_roc}, we also present the \emph{area under curve} (AUC) for each result because it represents the probability that a ranking method will score higher value on an arbitrary signal instance than the value of an arbitrary noise instance. For the computation of AUC, we use the \emph{trapezoidal} method.

\begin{figure*}[ht!]
{\includegraphics[width=.9\textwidth]{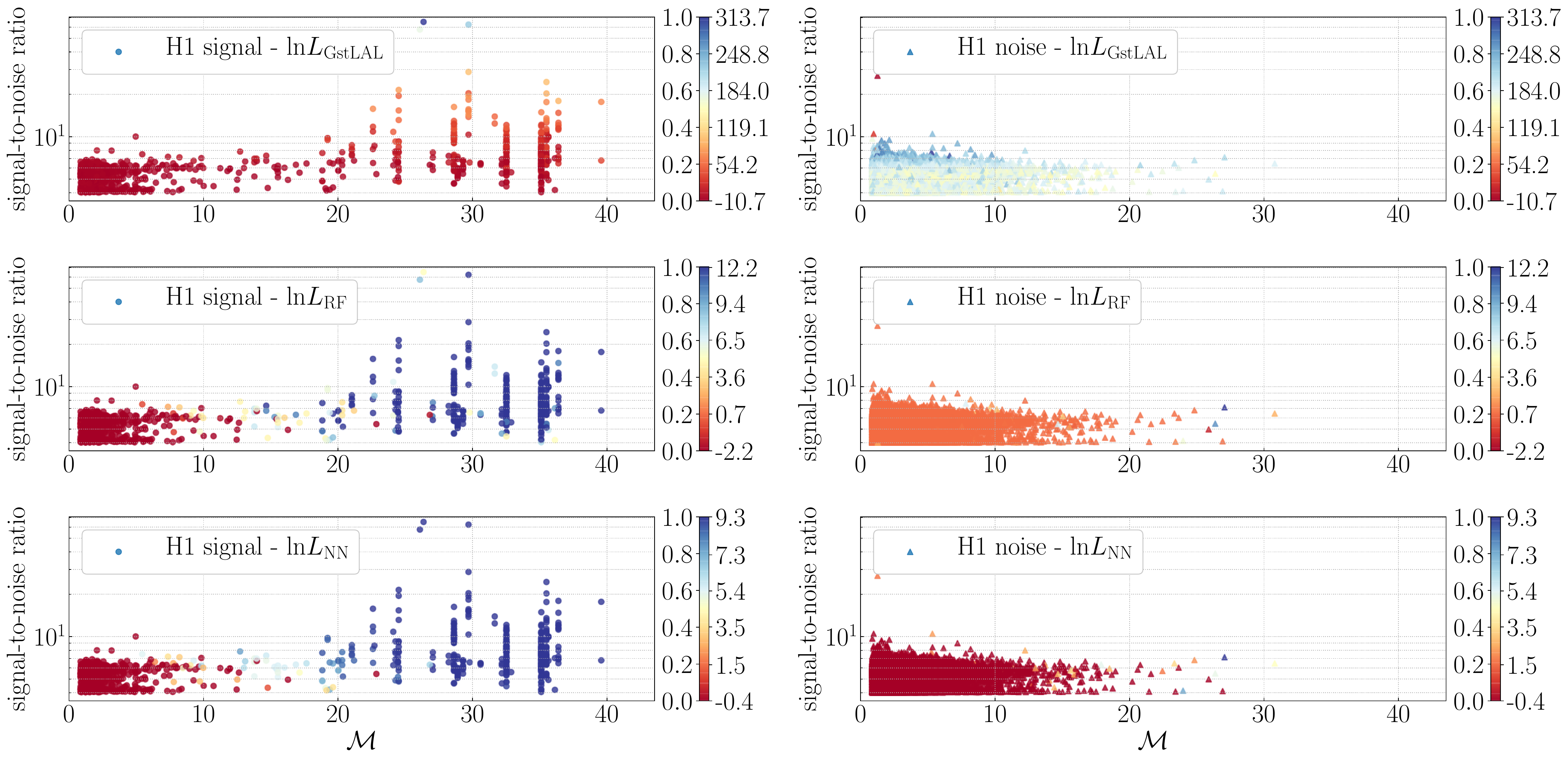}}
\caption{(Color online) Scatter plots of signal and noise samples of H1 data with respect to two selected feature parameters: SNR and the chirp mass, $\cal M$. The left and right columns show scatter plots of signal samples and noise samples, respectively. The color bar indicates the value of the normalized (left-side) and unnormalized (right-side) of $\ln L$s. One can see that MLs computed relatively higher $\ln L$ even for signal samples of lower SNR.}
\label{fig_scatter_h1}
\end{figure*}

From this figure, we can see that all cases are drawn in the upper region of the gray-dashed line which indicates random guess\footnote{The random guess is the case when a discriminator cannot distinguish a sample neither signal nor noise, i.e., the discriminator returns 0.5 for the probability of all signal and noise samples.}. We can understand this result as the ratio of signal samples having larger $\ln L$ is bigger than that of noise samples. In other word, we can say $\ln L^{(s)}_\mathrm{max}$ is bigger than $\ln L^{(n)}_\mathrm{max}$. From this result we know that $\ln L$ works as a proper ranking method in discriminating signal samples from noise samples as desired.

Also, one can see that two MLs show higher TPRs than GstLAL pipeline over given FPR ranges and it results in the AUCs of MLs are about 4--5\% larger than the AUCs of GstLAL pipeline. When we focus on the TPR at lower FPR region, in specific, at the lowest FPR where $\ln{L}^{}_{\textrm{th}}=\ln{L}^{(n)}_{\textrm{max}}$, MLs show {0.103--0.125} higher TPR than the TPR of GstLAL. This result shows that we could obtain highly probable signal samples {10.3\%--12.5\%} more with MLs than GstLAL pipeline. In the consideration of practical application of MLs to the low-latency search for GWs from binary mergers, the performance at lower FPR is important since FPR can be interpreted as the same analogy  to the false alarm probability, which will be discussed in Sec.~\ref{sec:sensitivity}, and, eventually, identifying a GW signal candidate with sufficiently low false alarm probability will be connected to the declaration of the detection of a GW signal.

\begin{figure*}[ht!]
{\includegraphics[width=.9\textwidth]{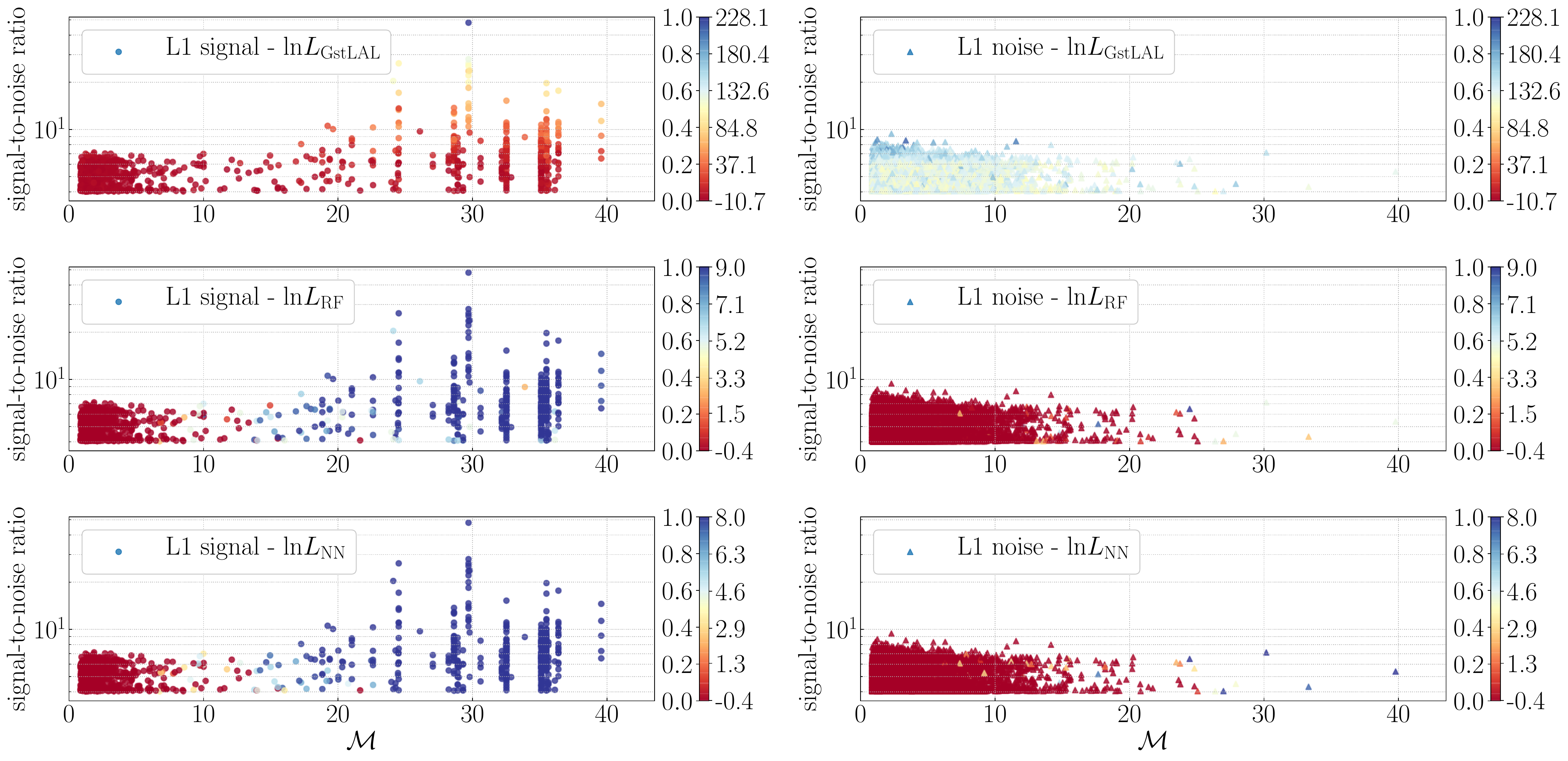}}
\caption{(Color online) Scatter plots of signal and noise samples of L1 data with respect to two selected feature parameters: SNR and the chirp mass, $\cal M$ as drawn in Fig.~\ref{fig_scatter_h1}. One can see that L1 data case shows similar with H1 data case but better discriminability on RF result than H1-RF result in Fig.~\ref{fig_scatter_h1}.}
\label{fig_scatter_l1}
\end{figure*}

In the comparison between RF and NN, one can notice that the performance of each ML is similar to each other, despite RF shows about 0.73--0.76\% larger AUC and about 1.1--1.7\% higher TPR at the minimum FPR than NN depend on data. From this result, one can recognize that the classification performance of RF on the overall test data is the best.

\subsection{Scatter Plot}
\label{sec:scatter_plot}

We need to examine how individual samples contribute to the resulted ROC curve. Thus, we investigate the contribution by looking the correlation between the value of $\ln L$s of considered ranking methods and the feature parameters. We present example scatter plots in Figs.~\ref{fig_scatter_h1} and~\ref{fig_scatter_l1} with color-bar showing normalized and unnormalized values of $\ln L$. Note that we use the chirp mass defined as 
\begin{equation}
{\cal M} = \frac{(m_1 m_2)^{3/5}}{ (m_1 + m_2)^{1/5}}
\end{equation}
instead of individual component masses, $m_1$ and $m_2$, for the horizontal axis because the chirp mass is one of important parameters in describing characteristics of the evolution of GW waveform generated from compact binary systems~\cite{jolien}. Also, since SNR is one of fundamental statistical quantities in judging the significance of a GW signal buried in noisy GW data, we especially select these two parameters for this example. 

From Figs.~\ref{fig_scatter_h1} and~\ref{fig_scatter_l1}, we see that signal samples of higher SNR and of higher chirp mass obtained higher $\ln L$s as expected from the ROC curve. In particular, for the distribution with respect to the chirp mass, we also see that signal samples in the range of $[4.5, 45] M_\odot$, which is believed as the detectable chirp mass range for BBH system by LIGO/Virgo. However, from the comparison of signal samples between GstLAL pipeline and MLs, GstLAL returns relatively lower $\ln L$s on samples having SNR $\lsim 10$ which is a criterion for SNR of GW candidate signal. On the contrary, MLs return higher $\ln L$s even for those signal samples. This result means MLs can distinguish even less significant signal samples correctly which may be disregarded as a candidate with GstLAL pipeline. 

On the other hand, for noise samples, all ranking methods returns relatively lower $\ln L$s than signal samples. However, in particular for H1 data, NN shows the best distinguishability than other two methods in terms of normalized $\ln L$. Meanwhile, for L1 data, RF and NN shows similar distinguishability. Therefore we conclude that the distinguishability on individual sample is less effective in the computation of TPR and FPR of ROC curve. We also see similar results from other scatter plots drawn with other feature parameters. One can find them from Appendix~\ref{sec:adx_scatter_plots}.


\section{Search Sensitivity}
\label{sec:sensitivity}

In this section, we discuss the search sensitivity through the relation between the sensitivity range and the \emph{false alarm rate} (FAR) \cite{far:2013prd} in order to suggest a practical application of MLs for searching GWs from compact binary systems. For this calculation, we refer the Sec.~IV.~C of Ref.~\cite{gstlal_inspiral:2017prd}.

In general, FAR is defined as
\begin{eqnarray}
\textrm{FAR} \equiv \frac{1}{T}{\int^{\infty}_{\ln L_{\mathrm{th}}} P(\ln L | n) d\ln L}, \label{eq_far}
\end{eqnarray}
where $T$ is the length of data segment. We can use FAR for the determination of a threshold value for a ranking method and, eventually, can use the value to judge a detection of a GW signal. If we assume that the chance of background noises gaining higher $\ln L$ is very rare, we can write the FAR for the given $\ln L$ of a noise in an approximated form:
\begin{eqnarray}
\textrm{FAR} \approx \frac{P(\ln L \geq {\ln L}_{\textrm{th}} | n)}{T}. \label{eq_far}
\end{eqnarray}
This approximation is valid for the consideration of this work too since it is shown that most of noise samples obtain lower $\ln L$ than signal samples.

The numerator of Eq.~(\ref{eq_far}) means the probability that noise data, $n$, get a high value of a ranking method. Thus it is called as \emph{false alarm probability} (FAP) \cite{far:2013prd}.
In this work, as discussed in Sec.~\ref{sec:roc}, FPR can be interpreted in the same analogy of FAP when we compare Eq.~(\ref{eq_fpr}) and the numerator of Eq.~(\ref{eq_far}). However, at this moment, we change the $\ln L_{\textrm{th}}$ in the expression of FAP to be $\ln L^{(s)}_\textrm{min} \leq \ln L_\textrm{th} \leq \ln L^{(s)}_\textrm{max}$ instead of using $\ln L^{(n)}_\textrm{min} \leq \ln L_\textrm{th} \leq \ln L^{(n)}_\textrm{max}$ since our interest is estimating the FAR of signal samples. 
Additionally, we follow the procedure for calculating Eq.~(31) of Ref.~\cite{gstlal_inspiral:2017prd} with the value of FPR in order to take account the corrections which have been using in the conventional GstLAL pipeline for the computation of FAP. 

For the calculation of the sensitivity range, we use the distance parameter which were used for the generation of the signal samples. We also adopt the definition of efficiency, $\epsilon$, with \emph{found} sample given in Ref.~\cite{gstlal_inspiral:2017prd}: when the estimated FARs of some signal samples are lower than a given fiducial FAR we call the signal samples as \emph{found} samples and the ratio of found samples to total number of signal samples as \emph{efficiency}. Then we compute the search volume such that
\begin{equation}
V = 4 \pi \int^\infty_0 \epsilon(r) l^2 dl \label{eq_search_volume}
\end{equation}
where $l$ is the distance to the source of GW and $\epsilon(l)$ is the efficiency at the distance $l$ and the sensitivity range, $R$:
\begin{equation}
R = \left( \frac{3V}{4 \pi} \right)^{1/3}.
\end{equation}

\begin{figure}[t!]
\includegraphics[width=.8\linewidth]{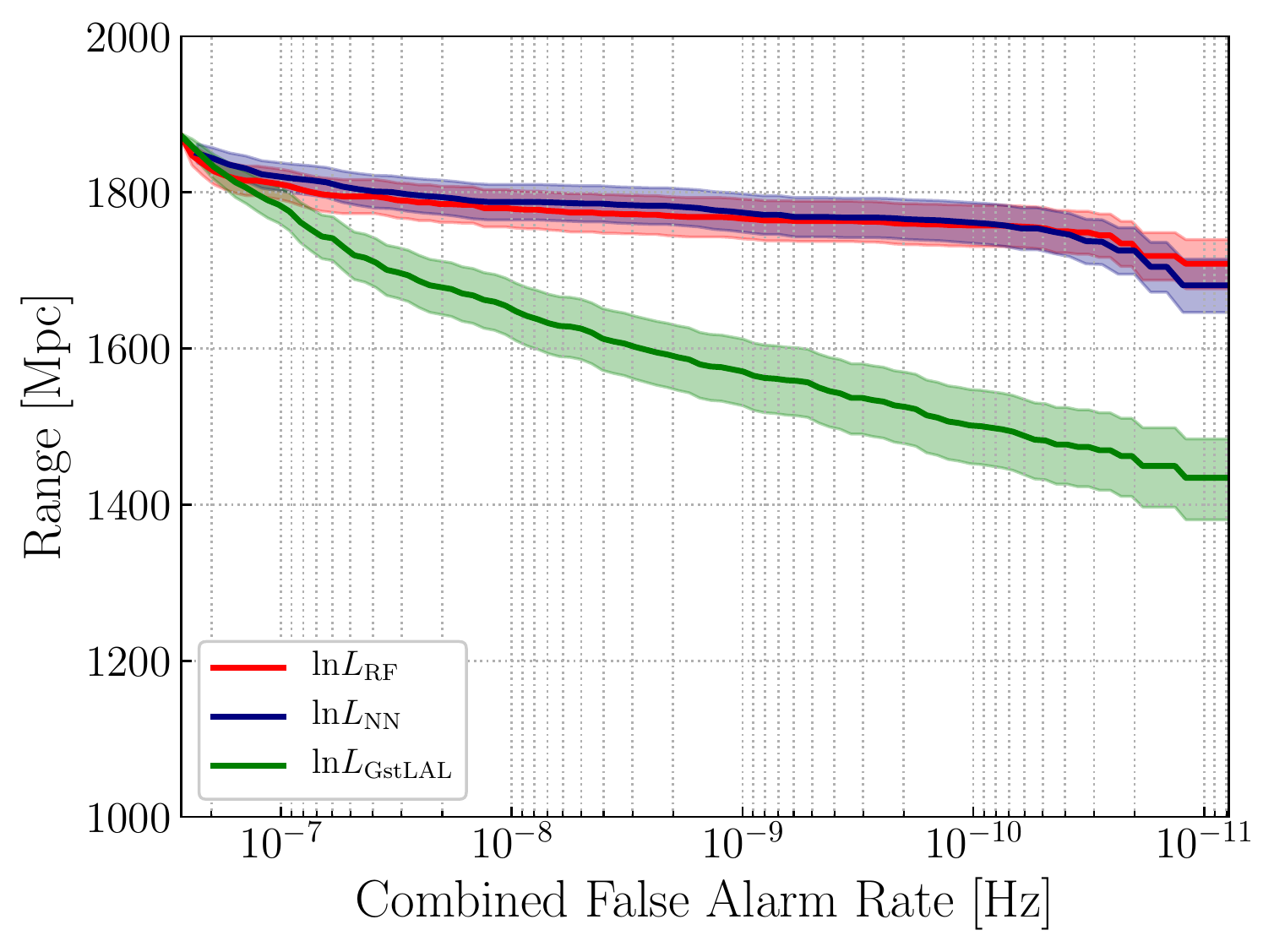}
\caption{(Color online) Comparisons of the sensitivity in terms of detectable range in distance versus the combined false alarm rate (FAR). The red line indicates the result of RF, the navy line indicates the result of NN, and the lime line indicates the result of GstLAL pipeline. The shaded regions around each line show $3\sigma$ of the binomial confidence interval computed based on Wilson method \cite{Wilson:1927} with continuity correction \cite{Newcombe:1998}. One can see that the detectable range of MLs is relatively farther than that of GstLAL pipeline at lower FAR region and it means we can identify an event occurred at more farther distance with MLs.
}
\label{fig_sensitivity}
\end{figure}

For the computation of the search volume, Eq.~(\ref{eq_search_volume}), we integrate the integrand with \emph{trapzoidal} method by varying the fiducial FAR. We refer the sensitivity range, $R$, as the search sensitivity in this work and plot it with respect to the combined FAR in Fig.~\ref{fig_sensitivity}. The combined FAR is obtained by collecting individual FAR computed with the data of each detector. In order to take account the uncertainty in binary discrimination into signal or noise, we compute the lower and upper bounds, $(\omega^-,\omega^+)$, of Wilson confidence interval \cite{Wilson:1927} with continuity correction \cite{Newcombe:1998}:
\begin{widetext}
\begin{eqnarray}
\omega^{-} &= \max& \left\{ 0, \frac{2{y}\hat{p} + z^2 - \left[ z \sqrt{ z^2 - 1/{y} + 4{y}\hat{p}(1-\hat{p}) + (4\hat{p}-2)} + 1 \right] }{2({y}+z^2)} \right\}, \label{omegam} \\
\omega^{+} &= \min& \left\{ 1, \frac{2{y}\hat{p} + z^2 + \left[ z \sqrt{ z^2 - 1/{y} + 4{y}\hat{p}(1-\hat{p}) + (4\hat{p}-2)} + 1 \right] }{2({y}+z^2)} \right\}, \label{omegap}
\end{eqnarray}
\end{widetext}
where $\hat{p} = p/{y}$ is the fraction of found signal samples to the number, $y$, of total samples and $z$ is the probit function. For Eqs.~(\ref{omegam}) and (\ref{omegap}),  if $\hat{p}=0$, $\omega^{-}$ is taken as 0. On the other hand, if $\hat{p}=1$, $\omega^{+}$ is taken as 1.

From Fig.~\ref{fig_sensitivity}, we can see that we can detect more farther events with MLs for given range of FAR than with GstLAL pipeline, in particular, at the lowest FAR, the central value of detectable range is $\sim$ 1.7 Gpc for both RF and NN while $\sim$ 1.4 Gpc for GstLAL pipeline. We see that the range of RF is slightly farther than that of NN. But, they are placed within $3\sigma$ uncertainty bounds of the Wilson confidence interval of each other. Therefore, we conclude their sensitivities are comparable and this result is consistent with the ROC curves discussed in Section \ref{sec:roc}.

\section{Summary and Discussion}
\label{sec:discussion}

Machine learning (ML) is known by its fast and accurate performance on identifying/classifying nonlinear multi-dimensional data of various fields. From several studies related to GW data analysis, applications of MLs have shown improved and/or comparable classification performances compared to conventional statistical approaches. Thus, in this work, we study the feasibility whether we can use the output of ML for ranking candidate events for low-latency GW searches. 

For this study, we consider two supervised MLs, random forest (RF) and neural network (NN). The mock data for GW150914 obtained by running the GstLAL pipeline in offline mode is used as the signal sample. From the output of GstLAL pipeline, we extract six physical and statistical feature parameters for the configuration of input data for ML. With given data, we train considered MLs and test the classification performance of the rank of MLs to compare it to the conventional ranking method, the log-likelihood ratio, $\ln L$ of GstLAL pipeline. However, MLs return only probabilistic rank values in between 0 and 1 while $\ln L$ of GstLAL pipeline is unnormalized. Thus, to make a fair comparison, we first estimate the probability density function (PDF) for ranks of signal and noise samples separately and then compute $\ln L$ with the PDF.

It is known that ML should be trained with sufficiently large and non-biased data for a successful application \cite{GeronMLbook}. In general, training a ML with a large data need long computational time from about hours to days to determine the most optimal combination for the hyperparameters of a ML. The training time depends on the size of train data such as the number of samples and the number of feature parameters. In this work, we find that training a ML with a set of train data of about 165,000 (number of samples) $\times$ 6 (number of features) takes a few hours. Meanwhile, evaluating an test data or a new data requires much shorter time than time for training: from our study, the evaluation with a set of test data of roughly 45,000 (number of samples) $\times$ 6 (number of features) dimensions can be done in the order of tens of milliseconds. Thus, we can see the positive prospect of applying ML for the low-latency GW search in terms of the analysis speed.

We investigate the classification performance of MLs through a couple of figure-of-merits. In this work, we choose the receiver operating characteristic (ROC) curve to see overall performance for all tested samples and the scatter plot to see the contribution to individual sample. From the ROC curve and the area under curve, we can observe that MLs show better performance on classifying more signal samples from noise samples than GstLAL pipeline. The result on individual samples is also studied through the scatter plot to try to see the correlation between the output value of a given ranking method and selected feature parameters, e.g., signal-to-noise ratio (SNR) and chirp mass. We find that signal samples of relatively higher $\ln L$s have higher SNR and chirp masses within expected chirp mass range for binary black hole mergers. Thus we are convinced that the resulted higher $\ln L$ values of given ranking methods are correlated to the tested feature parameters and could see similar correlation from other scatter plots in Appendix \ref{sec:adx_scatter_plots}.

For the difference in performance between RF and NN, one may suspect that the hyperparameters of NN might be less optimized than RF since we empirically selected the hyperparameters of NN without automated determination as discussed in Appendix~\ref{sec:adx_nn_hyp}. However the most optimal choice on the hyperparameter depends on the input data for training: if we train MLs with different training data, the choice on the set of optimal hyperparameters will be also changed. On top of that, it is hard to think the data used in this work can represent the general property of all possible BBH systems. Thus, conducting more fine-tuning on the optimal hyperparameters for NN with the data used in this work is out of scope of this kind of feasibility study and we admit the difference between RF and NN is placed in acceptable range.

We compare the sensitivity in terms of the detectable range with respect to the approximated false alarm rate (FAR) too. In specific, since the false positive rate (FPR) of ROC curve can be translated to the false alarm probability, the FPR of ROC curve is also used in computing the approximated FAR. From the sensitivity plot, we can see that MLs are more sensitive than the GstLAL pipeline, that is, it would be possible to detect farther events with MLs beyond the upper limit of the detectable range of the GstLAL pipeline. Therefore we conclude that using output of ML can be an alternative ranking method to the conventional ranking method of GstLAL pipeline and it is worth to consider ML as a new ranking method for future low-latency searches for GWs from binary mergers.

In this work, we constrained the origin of feature parameters of input data for ML only to the information obtained from the GstLAL pipeline for simplicity. However, it is also possible to consider to collect transient noise informations, which are used for the GW data quality measurement, along with the current methodology. In the future work, therefore, we will discuss about the practical implementations such as training MLs with combining the current feature parameters and transient noise information into the input data. Next, we will build up the strategy and the framework for an online GW inspiral search pipeline which implementing ML as its ranking method.

However we admit that this approach is rather weak in interpretability: it is not easy to clearly understand how the model results in the better result with less information than the conventional method. This point is one of differences from the conventional approach because it is built on statistically reliable considerations and, eventually, has strong interpretability. The interpretability is another critical point for judging the confidence of a detection. Therefore, for the practical implementation, we may also need to design an additional method to make the classification model to be interpretable.


\begin{acknowledgments}
The authors thank the LIGO Scientific Collaboration for the use of the mock data for GW150914 and corresponding background data. KK would like to specially thank to J.~J.~Oh, S.~H.~Oh, E.~J.~Son, Y.~-M.~Kim, W.~Kim, J.~Lee, S.~Caudill, and K.~Cannon for fruitful and constructive discussions. The work described in this paper was partially supported by grants from the Research Grants Council of the Hong Kong (Project No. CUHK 14310816 and CUHK 24304317) and the Direct Grant for Research from the Research Committee of the Chinese University of Hong Kong. \end{acknowledgments}


\appendix

\section{Hyperparameters of RF}
\label{sec:adx_rf_hyp}

For RF, we run a module, \texttt{GridSearchCV} embedded in Scikit-Learn for searching optimal hyperparameters. The core of this module is the \textit{$k$-fold cross validation} method: conducting $k$-times validation tests on $k$ different validation subsets, which are prepared by shuffling all train data and then dividing evenly into $k$ subsets, with a given combination of hyperparameters. Each validation test is done by evaluating one of $k$ subsets as a test subset based on the trained ML which is trained with remaining subsets. This module computes the \emph{averaged accuracy}:
\begin{equation}
\textrm{avg. accuracy} \equiv \frac{\textrm{number of samples}({y_{\textrm{true}} = {y_{\textrm{pred}}}})}{\textrm{total number of samples in a subset}} \label{eq_avg_acc}
\end{equation}
as the final output of each test for RF. In Eq.~(\ref{eq_avg_acc}), $y_{\textrm{true}}$ and $y_{\textrm{pred}}$ denote, respectively, the original class and the predicted class of a sample. At last, a combination of hyperparameters which gives the highest averaged accuracy is selected as the most optimal set of hyperparameters. Used entries of selected hyperparameters for running this module are tabulated in Table \ref{gridsearch_param}. One can find the description of each hyperparameter from Ref.~\cite{hyperparam_rf}. 

\begin{table}[h!] 
\caption{Tested entries for hyperparameters of RF in running \texttt{GridSearchCV}.}
\begin{center}
\begin{tabularx}{\columnwidth}{X  X}
\hline
\hline
{Hyperparameter} & {Entry} \\
\hline
\texttt{n\_estimators} & {50, 100, 200} \\
\hline
\texttt{criterion} & {gini, entropy} \\
\hline
\texttt{max\_features} & {2, 4, 6} \\
\hline
\texttt{min\_samples\_split} & {2, 3, 4, 5} \\
\hline
\texttt{max\_depth} & {None, 10, 30, 50} \\
\hline
\hline
\end{tabularx}
\end{center}
\label{gridsearch_param}
\end{table}

In this work, we conduct the run of \texttt{GridSearchCV} with $k=3$ and the determined optimal hyperparameters for given data are summarized in Table \ref{tab:optimal_param}. From this table, one can see that some hyperparameters are the same for different data. It means those values are the most optimal value amongst tested entries for the type of data of this work. Therefore, if we do not change the selected six feature parameters for similar type of data, we can fix the values of those hyperparameters and alter remains for the determination of optimal hyperparameters.

\begin{table}[t!] 
\caption{Empirically determined optimal hyperparameters of RF by running \texttt{GridSearchCV}. One can see that some hyperparameters are the same for different data.}
\begin{center}
\begin{tabularx}{0.99\linewidth}{l  X  X}
\hline
\hline
{Hyperparameter} & {Data} & {Optimal} \\
\hline
\multirow{2}{*}{\texttt{n\_estimators}} & H1 & {50} \\
\cline{2-3}
& L1 & {50} \\
\cline{1-3}
\multirow{2}{*}{\texttt{criterion}} & H1 & {entropy} \\
\cline{2-3}
& L1 & {entropy}\\
\cline{1-3}
\multirow{2}{*}{\texttt{max\_features}} & H1 & {4} \\
\cline{2-3}
& L1 & {4} \\
\cline{1-3}
\multirow{2}{*}{\texttt{min\_samples\_split}} & H1 & {3} \\
\cline{2-3}
& L1 & {4} \\
\cline{1-3}
\multirow{2}{*}{\texttt{max\_depth}} & H1 & {30} \\
\cline{2-3}
& L1 & {10} \\
\hline
\hline
\end{tabularx}
\end{center}
\label{tab:optimal_param}
\end{table}

\begin{table}[h!] 
\caption{Hyperparameters for NN.}
\begin{center}
\begin{tabularx}{\columnwidth}{X  X}
\hline
\hline
{Hyperparameter} & {Value} \\
\hline
\multirow{3}{*}{layers (nodes)} & 1 input (6 nodes) \\ \cline{2-2}
& 4 hidden (32 nodes for each) \\ \cline{2-2}
& 1 output (1 node) \\
\hline
learning rate & {0.01 \%} \\
\hline
regularization & {$L2$ with 0.01 \%} \\
\hline
dropout & {10 \%} \\
\hline
activation function & {ReLU} \\
\hline
cost function & Cross entropy with Softmax \\
\hline
batch size & 1024 \\
\hline
\hline
\end{tabularx}
\end{center}
\label{NN_hyperparam}
\end{table}

\section{Hyperparameters of NN}
\label{sec:adx_nn_hyp}

Unlike RF, we set the hyperparameters of NN empirically because there is no available module for grid search in TensorFlow.\footnote{It is also known that it is hard to consider such grid search model for DNN because of there are too many hyperparameters to be tuned~\cite{GeronMLbook}. However, it is not impossible at all if we use an automated method such as DeepHyper~\cite{deephyper}. Despite of the availability of automated hyperparameter search method, we do not implement it in this work because we could get satisfactory performance with the empirically determined hyperparameters.} We set them to be the same for all considered data for convenience. In addition to the hyperparameters for the topology of a NN, in order to avoid overestimation (or overfitting equivalently), we adopt $L2$ \textit{regularization} to constrain a NN's connection weights and \textit{dropout} \cite{dropout:arXiv2012} to remove potential dependency on certain nodes of given network. The \textit{rectified linear units} (ReLU) function \cite{relu:2010} is used for the activation function between nodes in two adjacent layers. For the cost function, which measures the error between the target value, 1 for signal and 0 for background, and the output value in between 0 and 1 of the output node, \textit{Cross Entropy} function is implemented by taking the output probability computed from \textit{Softmax} function as the input probability of the Cross Entropy function. Finally, we also consider \textit{Batch Normalization} \cite{batch_norm:arXiv2015} to properly minimize the cost function in the \textit{backpropagation} \cite{backpropagation} process. All of these hyperparameters are summarized in Table \ref{NN_hyperparam} as well.

\section{Probability Density Estimation}
\label{sec:adx_pdf}

\begin{figure}[h!]
\subfloat[H1 data with RF]{\includegraphics[width=0.45\linewidth]{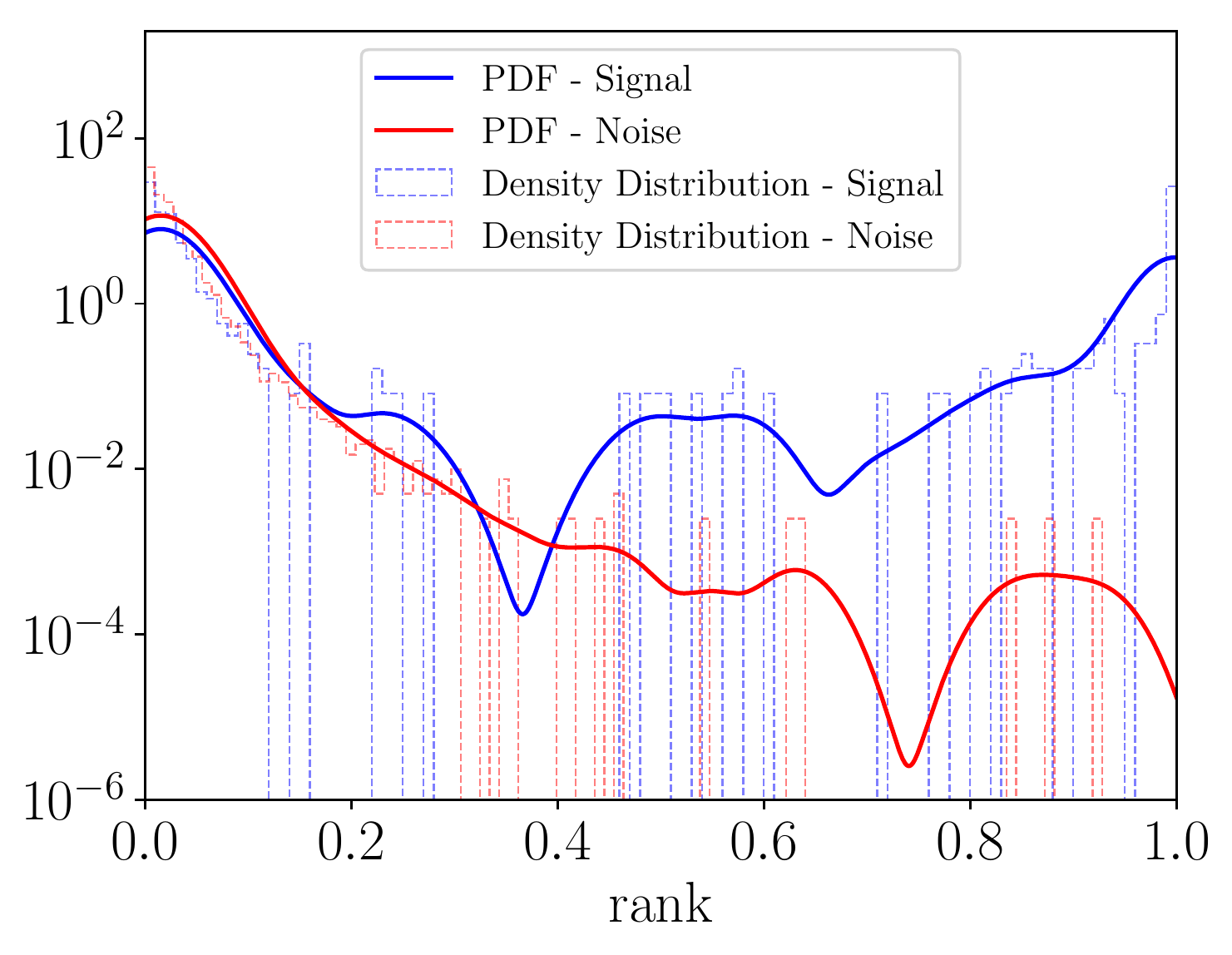}}
\subfloat[H1 data with NN]{\includegraphics[width=0.45\linewidth]{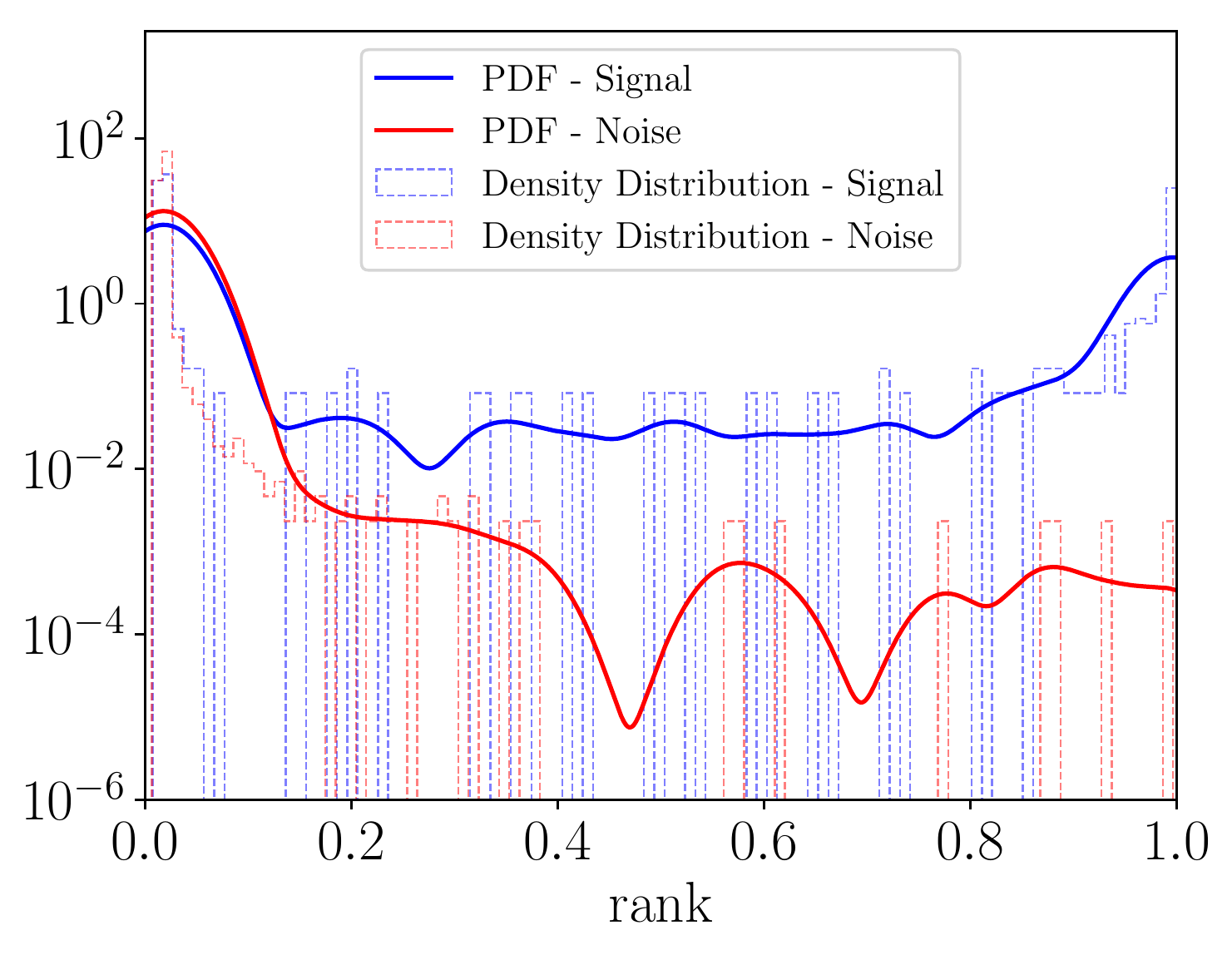}} \\
\subfloat[L1 data with RF]{\includegraphics[width=0.45\linewidth]{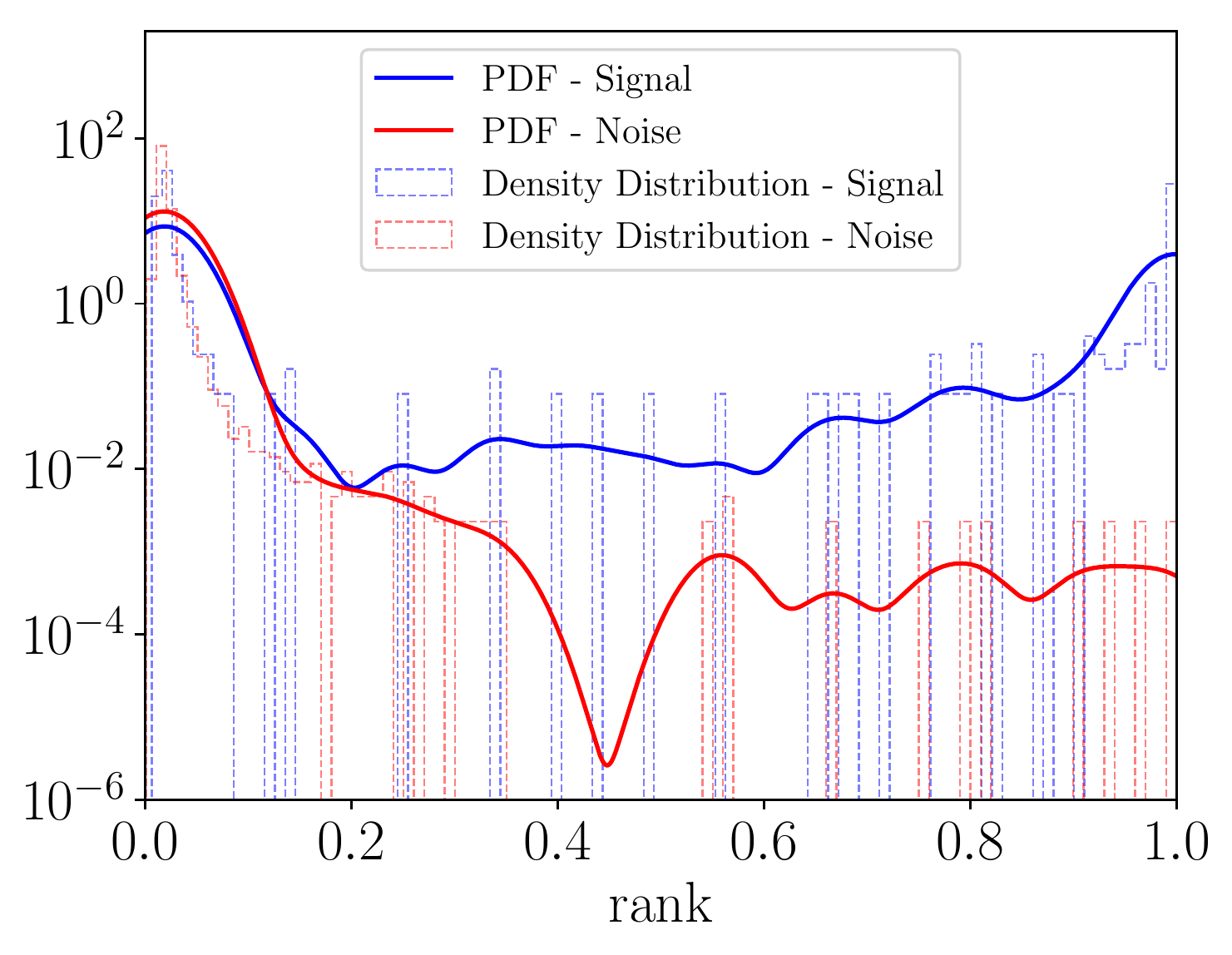}}
\subfloat[L1 data with NN]{\includegraphics[width=0.45\linewidth]{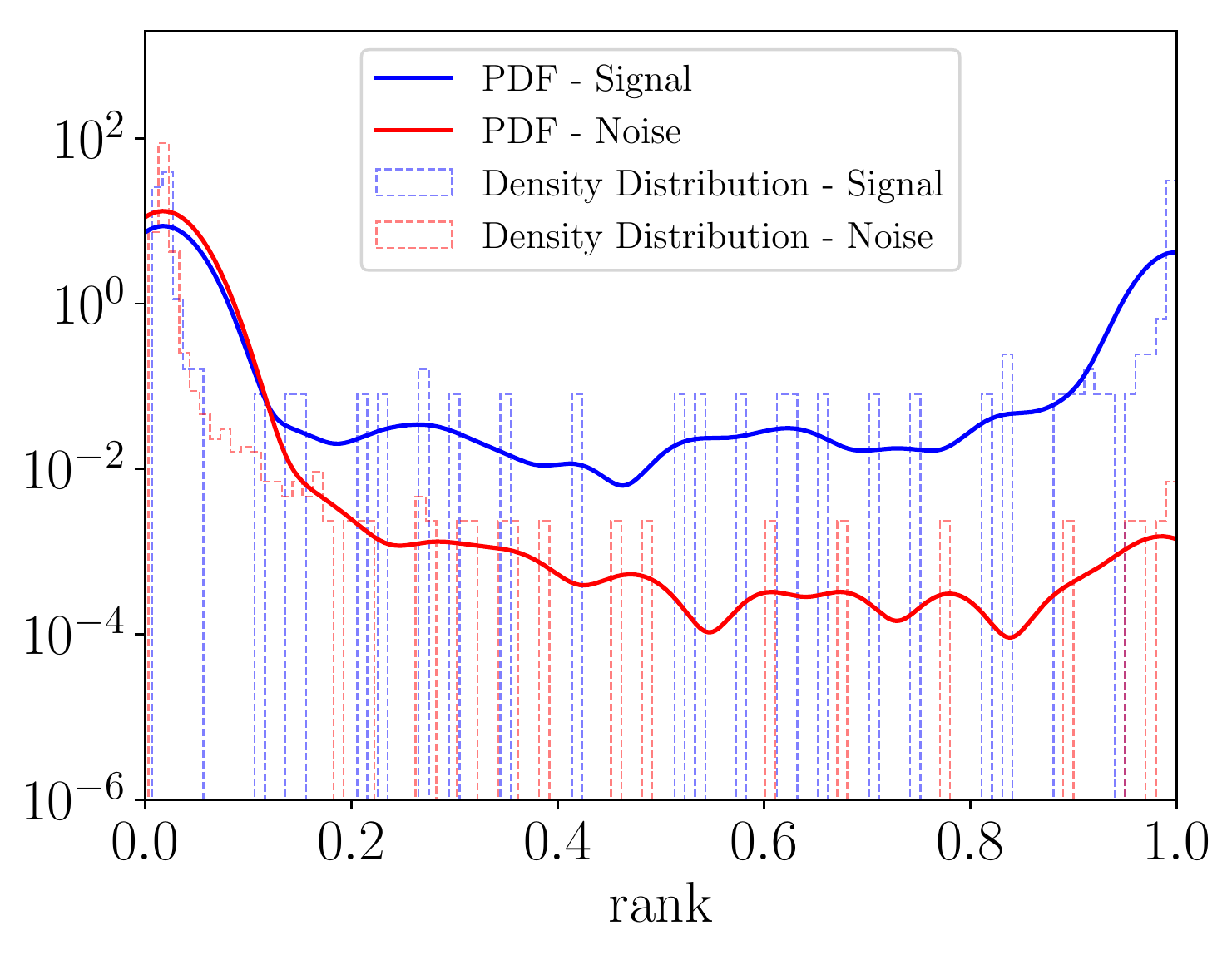}} 
\caption{(Color online) Estimated probability density functions (PDFs) for each data with studied MLs. For the PDF estimation, we use Kernel Density Estimation method of Scikit-Learn with Guassian kernel and bandwidth of 0.03.  The blue- and red-solid line are the estimated PDFs for signal samples and for noise samples, respectively. The blue- and red-dashed boxes show the normalized density distribution of signal and noise samples, respectively, used for the estimation of PDFs.}
\label{fig_pdf}
\end{figure}

In order to compute log-likelihood ratio, we estimate the probability density functions (PDFs) of signal and noise samples for numerator and denominator respectively of Eq.~(\ref{eq:lnL_ML}). The estimation of PDFs of resulted ranks of evaluation samples is done by using the \emph{Kernel Density Estimation} (KDE) method of Scikit-Learn with Gaussian kernel and an empirically determined optimal bandwidth of $0.03$. For this estimation, we compute normalized density distribution for the ranks of evaluation samples first and apply KDE with given bandwidth. The resulted PDFs for signal and noise samples of each test data using ranks from each ML are shown in Fig.~\ref{fig_pdf}. From these plots and Eq.~(\ref{eq:lnL_ML}), we can expect that we can obtain smaller $\ln L$ with lower ranks and bigger $\ln L$ with higher ranks.

\begin{figure*}[t!]
\subfloat[H1 data]{\includegraphics[width=0.9\textwidth]{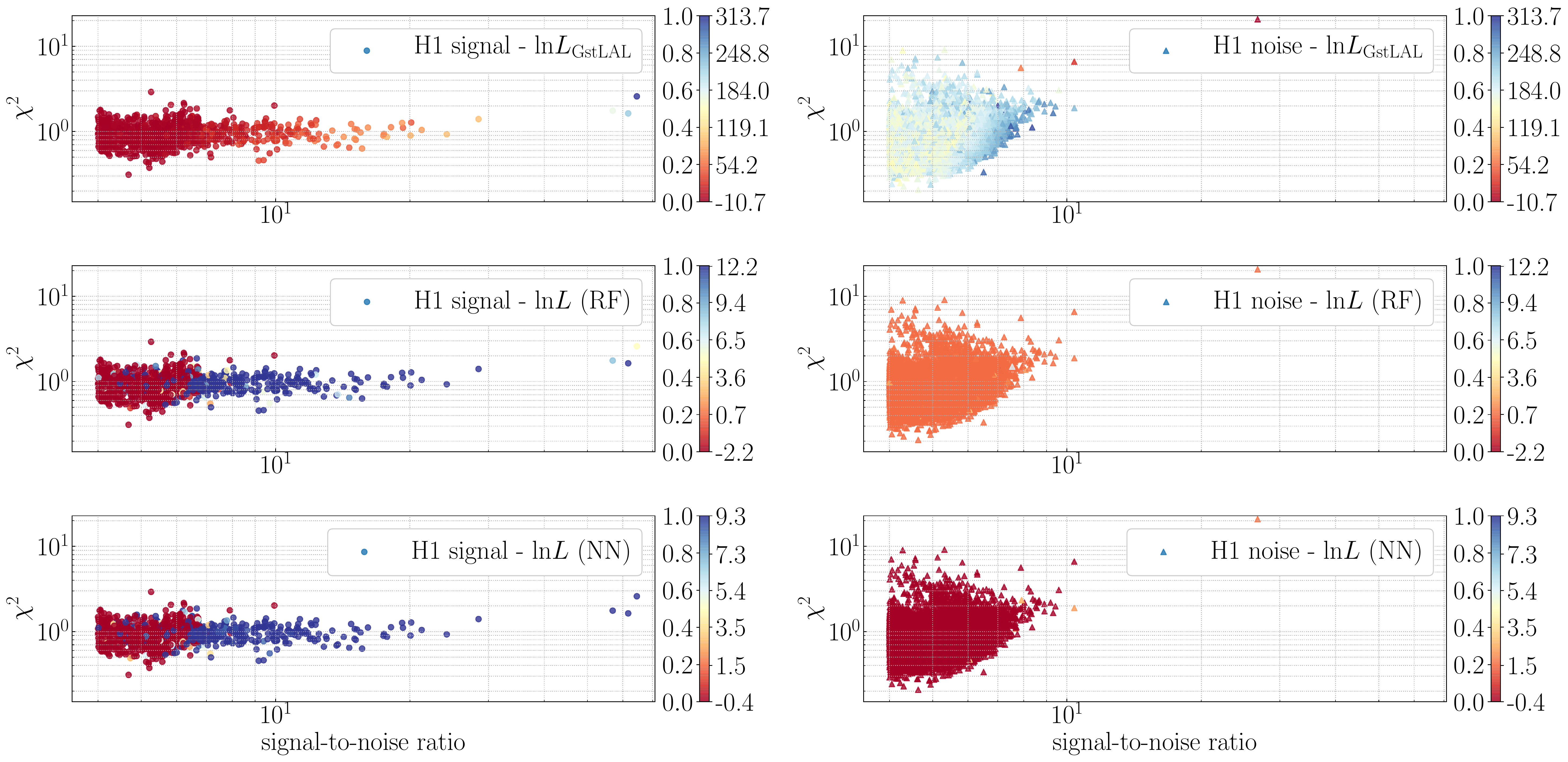}} \\
\subfloat[L1 data]{\includegraphics[width=0.9\textwidth]{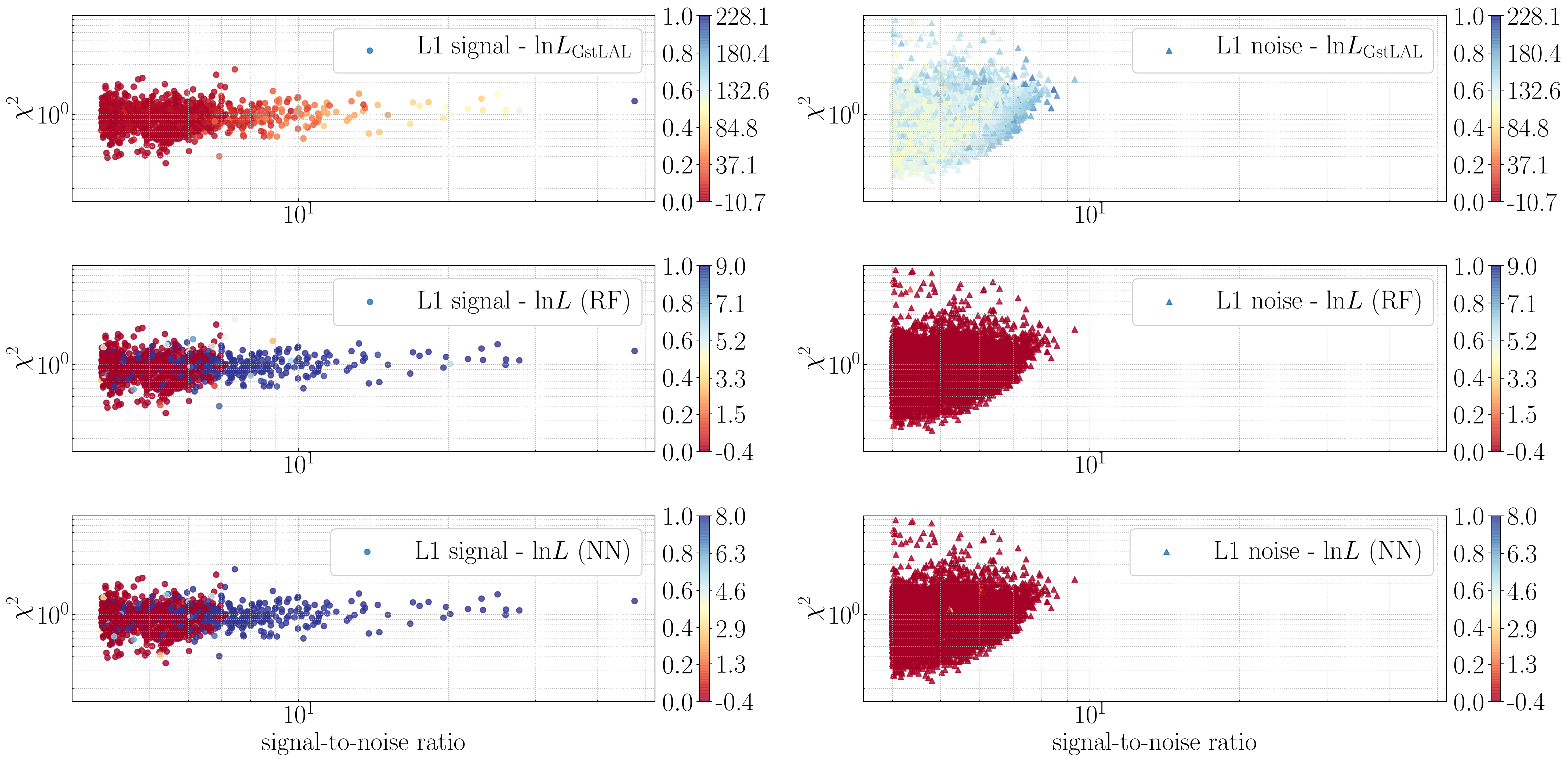}}
\caption{(Color online) Scatter plots of two selected feature parameters, SNR and $\chi^2$. }
\label{fig_scatter_snr_chisq}
\end{figure*}
\section{Scatter Plots}
\label{sec:adx_scatter_plots}

\begin{figure*}[]
\subfloat[H1 data]{\includegraphics[width=0.9\textwidth]{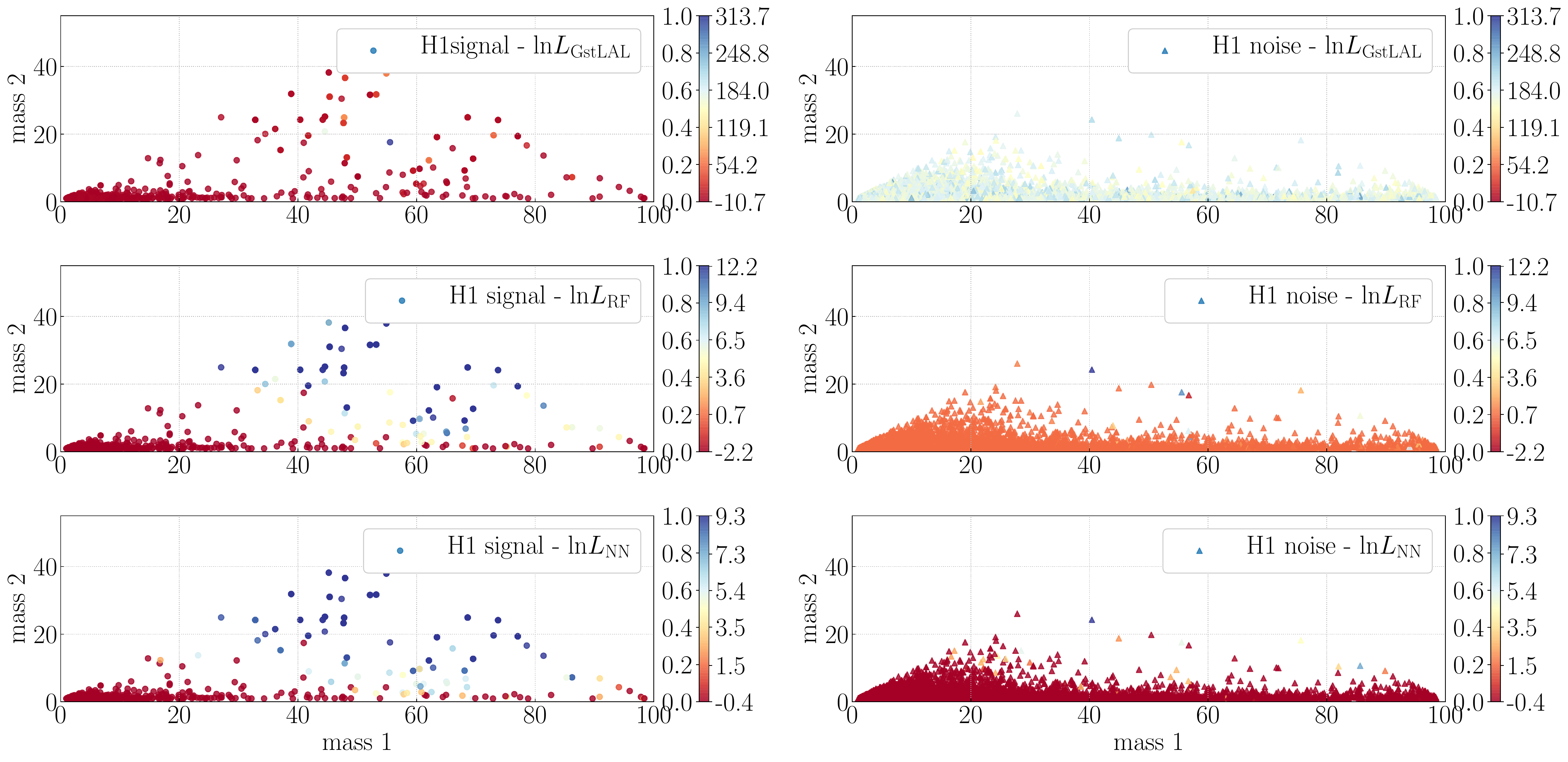}} \\
\subfloat[L1 data]{\includegraphics[width=0.9\textwidth]{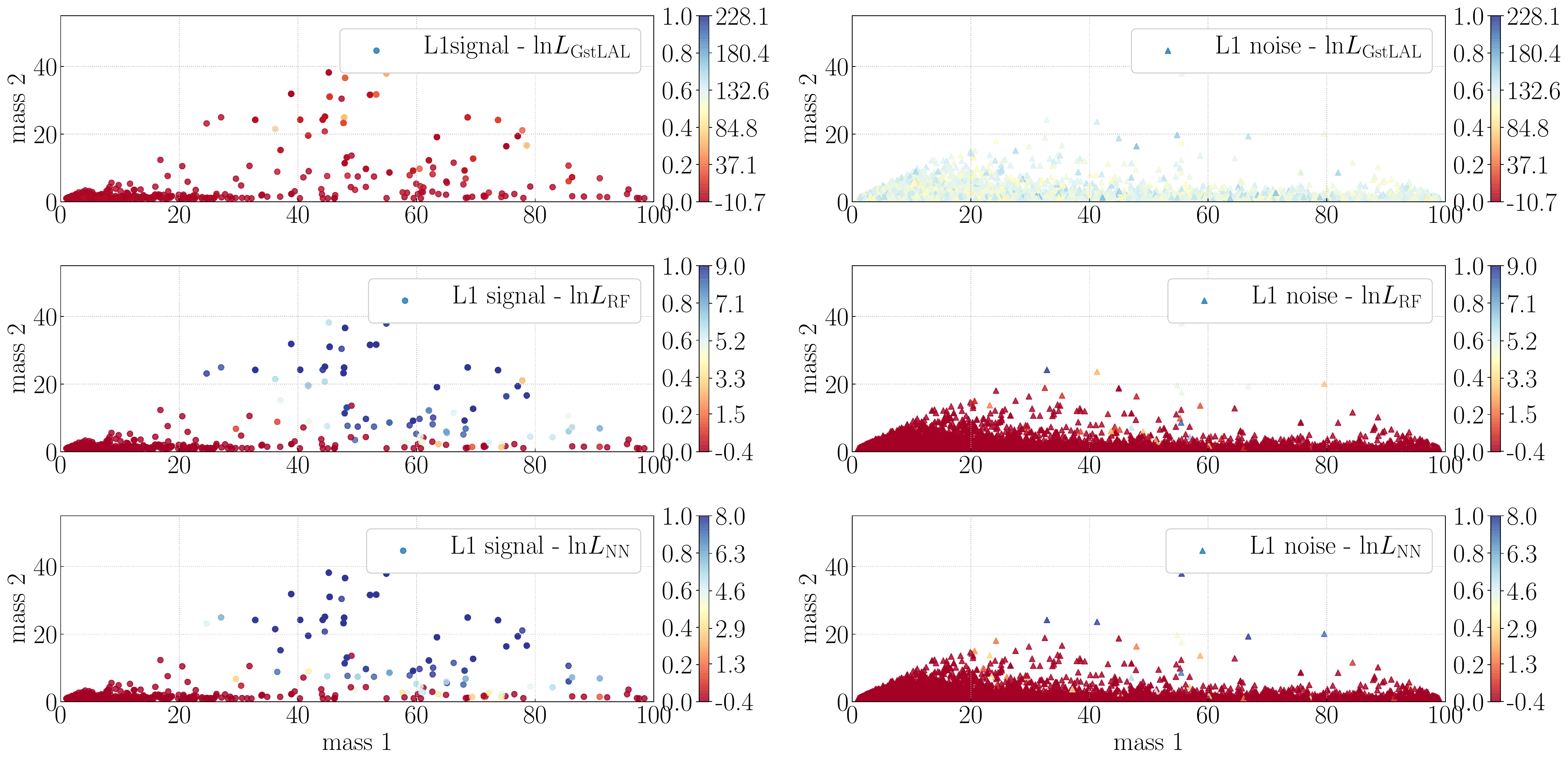}}
\caption{(Color online) Scatter plots of two selected feature parameters, mass 1 and mass 2. }
\label{fig_scatter_m1_m2}
\end{figure*}

\begin{figure*}[]
\subfloat[H1 data]{\includegraphics[width=0.9\textwidth]{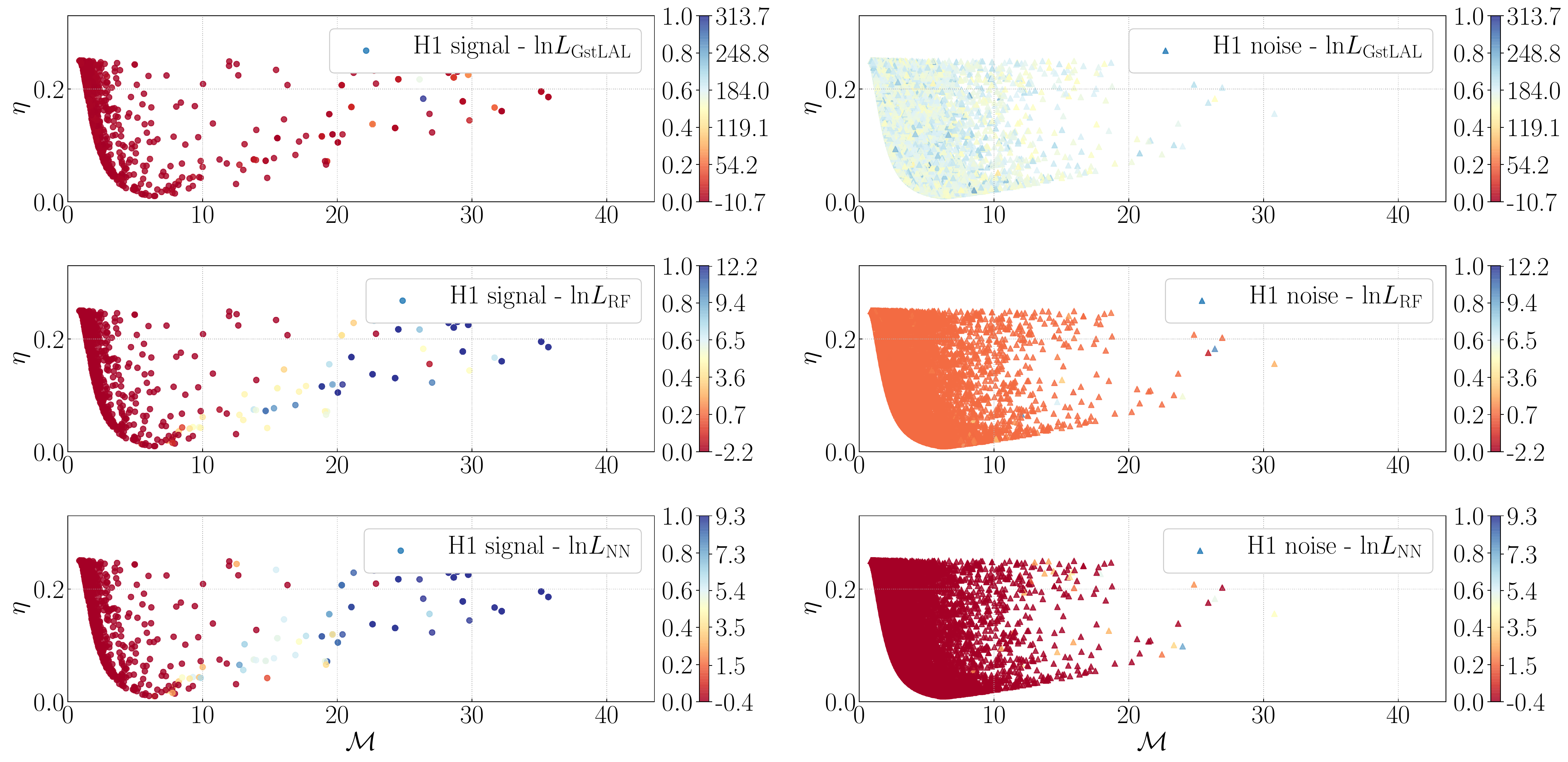}} \\
\subfloat[L1 data]{\includegraphics[width=0.9\textwidth]{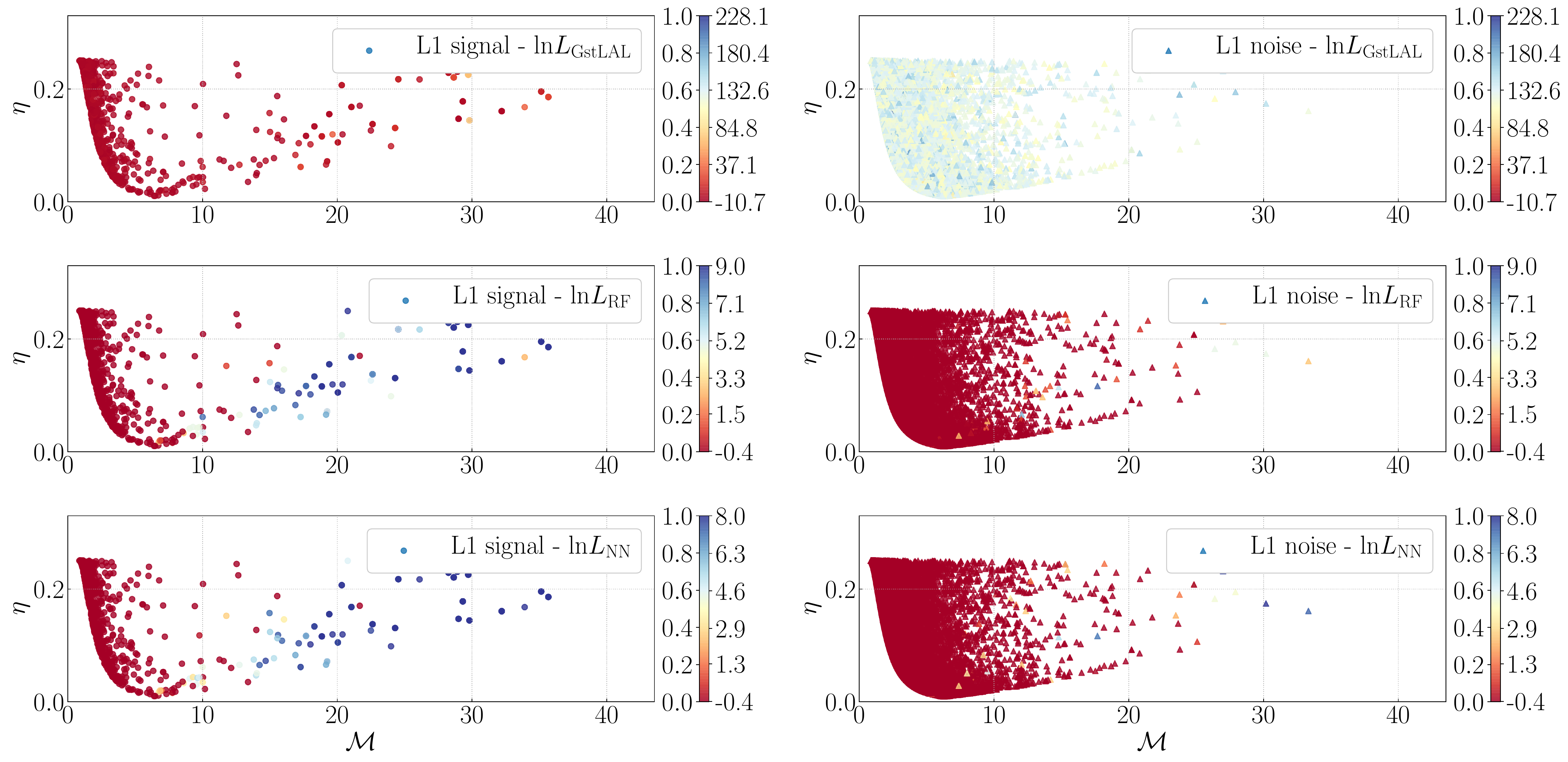}}
\caption{(Color online) Scatter plots of two selected feature parameters, $\cal M$ and $\eta$. }
\label{fig_scatter_eta_mc}
\end{figure*}

In this section, we present other scatter plots (Figs.~\ref{fig_scatter_snr_chisq}, \ref{fig_scatter_m1_m2}, and \ref{fig_scatter_eta_mc}) of signal samples and noise samples drawn with respect to several feature parameters, SNR, $\chi^2$, and component masses, $m_1$ and $m_2$ except those plots presented in Sec.~\ref{sec:scatter_plot}. In addition to these parameters, we also present scatter plots of the \emph{symmetric mass ratio}, $\eta \equiv \mu / M$, and the chirp mass, $\cal M$, where $\mu$ is the reduced mass and $M$ is the total mass, that is, $M = m_1 + m_2$. From all figures, we can see MLs returns relatively higher $\ln L$s on more signal samples as GstLAL pipieline does for signal-like samples, i.e., higher SNR, lower $\chi^2$, and adequate mass range for BBH merger. But, we can see that MLs can score higher $\ln L$ even for less significant signal samples, e.g., samples of SNR $\lsim 10$. Also, MLs show better distinguishability between signal and noise samples than GstLAL. These results are consistent with the result of ROC curve. Thus we advocate using output of MLs to compute $\ln L$ for ranking GW candidate events is promising approach.


\clearpage
\bibliography{MLtoGstLALbib}

\end{document}